\documentstyle[epsf,preprint]{jpsj}
\title{The impact of two-dimensional elastic disk}
\author{Hiroto {\sc Kuninaka}\footnote{E-mail: kuninaka@yuragi.jinkan.kyoto-u.ac.jp} and Hisao {\sc Hayakawa}\footnote{E-mail: hisao@yuragi.jinkan.kyoto-u.ac.jp}}
\inst{Graduate School of Human and Environmental Studies,
 Kyoto University, Sakyo-ku, Kyoto 606-8501}
\recdate{\today}
\kword{impact, elastic disk, coefficient of restitution}
\begin{document}
\sloppy
\maketitle

The impact of macroscopic materials is 
characterized by the coefficient of restitution (COR)
defined by
\begin{equation}
e=-v_{r}/v_{i} ,
\end{equation}
where $v_i$ and $v_r$ are the relative velocities of incoming and outgoing
 particles, respectively. 
COR $e$ had been believed to be a material constant.
 In general, however, experiments show that COR
 for three dimensional materials 
is not a constant even in approximate sense
 but
 depends strongly on
 the impact velocity\cite{goldsmith,sonder,bridges,basile}.

The origin of the dissipation in inelastic collisions is the transfer of
the kinetic energy of the center of mass 
into the internal degrees of freedom during the impacts.
 Gerl and Zippelius\cite{gerl} performed the
 microscopic simulation of two-dimensional collision of an elastic disk
 with a wall.  Their simulation is mainly based on the mode expansion of 
an elastic disk under the force free boundary condition. Then, they
 analyze Hamilton's equation ;
\begin{equation}\label{H's equation}
\dot P_{n,l}=-\frac{\partial H}{\partial Q_{n,l}} ;\quad
\dot Q_{n,l}=\frac{\partial H}{\partial P_{n,l}}
\end{equation}
under the Hamiltonian
\begin{equation}\label{hamiltonian}
H=\frac{p_0^2}{2M}+\sum_{n,l}^N(\frac{P_{n,l}^2}{2M}+\frac{1}{2}M\omega_{n,l}^2
Q_{n,l}^2)+V_0\int_{-\pi/2}^{\pi/2}d\phi \textup{e}^{-ay(\phi,t)} .
\end{equation}
Here $Q_{n,l}$ and $P_{n,l}$ are respectively the expansion coefficient 
of the elastic deformation and the canonical momentum, where $n$ and $l$
are the mode indices. $y(\phi,t)$ is the shape of the elastic disk in the polar
coordinate\cite{gerl,neptis}. $M$ is the mass of the disk, and 
$p_0$ satisfies 
$ \dot p_0=M \ddot y_0=- (\partial H/\partial y_0)$
with the position of the center of mass $y_0$.
 $V_0$ and $a$ are parameters to express
 the strength of the wall potential and
 $\omega_{n,l}$ is the angular frequency of the $(n,l)$ 
mode\cite{gerl,neptis}. 
Their results indicate
 that COR decreases with the impact velocity, which
strongly depends on Poisson's ratio. 
Since the relation
 between 
 quasi-static theory of impact\cite{kuwabara,brilliantov96,morgado} 
and their microscopic
 simulation\cite{gerl} is not clear,
we have to clarify the relation between two typical approaches.

In this short note, we will perform the microscopic simulation of the impact 
of a two-dimensional elastic disk with a wall. We introduce
a continuum model which is identical to that by Gerl and
Zippelius\cite{gerl}. Through our simulation,
 we found that this model does not recover
 the quasi-static theories\cite{kuwabara,brilliantov96,morgado}.
Details of this short note have been published
 as a longer conference report.\cite{neptis}      

Let us explain our model. In this model, the 
wall exists at $y=0$, and the center of mass keeps the
position at $x=0$. The disk approaches from the region $y > 0$ and is rebounded by the wall.
As in the simulation by Gerl and Zippelius\cite{gerl} we introduce the
 wall whose potential is  
given by
$V_{0}\textup{e}^{-ay}$
in this model,
 $V_{0}=aMc^{2}/2$ and $a=500/R$ with the radius of the disk $R$.
 This choice of $a$ is aimed to simulate the collision between two
 identical disks, though we have not extrapolated our
 results to the limit of $a \to \infty$.
 Such the extrapolation has been checked by Gerl and Zippelius for
 this model\cite{gerl}.
 We only simulate the case of Poisson's ratio $\sigma=1/3$.
 The numerical scheme of the integration of this model is
 the fourth order symplectic integral method
 with $\Delta t=5.0\times 10^{-3}R/c$ with $c=\sqrt{Y/\rho}$
where $Y$ is Young's modulus and $\rho$ is the density.

\begin{figure}[htbp]
 \epsfxsize=10cm
 \centerline{\epsfbox{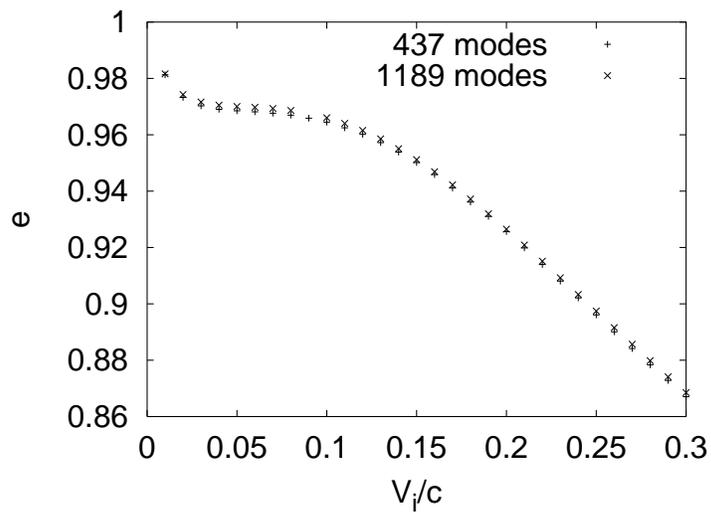}} 
 \caption{Authors: Hiroto Kuninaka and Hisao Hayakawa}
 \label{rest}
\end{figure}

At first, we carry out the simulation of this model  
with the initial condition at $T=0$ (\textup {i.e.} no internal
motion).
Figure \ref{rest} is the plot of the COR against the impact velocity. 
 We show the results of $437$ modes
 and $1189$ modes\cite{neptis}
 which clearly demonstrates the convergence
 of the result for the number of modes.
 Each line decreases smoothly as impact velocity increases.

Second,
 we investigate the force acting on the center of mass of the disk
caused
 by the interaction with the wall in this model.
In the limit of $v_i\to 0$, we expect that the Hertzian contact 
theory can be used\cite{landau,johnson,gerl}.
The small amount of transfer from the translational motion
 to the internal motion is the macroscopic dissipation.
Thus, we  can check the validity of quasi-static 
approaches\cite{kuwabara,brilliantov96,morgado}
 from our simulation by the difference between the observed force acting on
 the center of mass and the Hertzian contact force.
 The two dimensional Hertzian contact law\cite{johnson,gerl} is given by
the relation between the macroscopic deformation of the center of mass
 $h$ and the elastic force $F_{el}$  as
\begin{equation}
h\simeq-\frac{F_{el}}{\pi Y}\{ \ln \left(\frac{4 \pi Y
        R}{F_{el}\left(1-\sigma^{2}\right)}\right)-1\} ,\label{eq:hertz}
\end{equation}
where $Y$, $\sigma$ and $R$ are the Young modulus, Poisson's ratio and
the radius of the disk without deformation, respectively.
If $h$ is given, we can calculate the
 elastic force by solving eq.(\ref{eq:hertz}) numerically.
 Since in the limit of $v_i\to 0$  we may replace eq.(\ref{eq:hertz})
by $F_{el}\simeq -\pi Y h/\ln(4R/h)$\cite{gerl}. 
 Thus, the dissipative force $F_{dis}$
 in the two-dimensional quasi-static theory
are expected to be $F_{dis}\propto{-\pi Y \dot h}/{\ln(4R/h)}$.
 The total force $F_{tot}$ in the two-dimensional quasi-static theory
 are expected to be
\begin{equation}
 F_{tot}\simeq -\frac{\pi Y  h}{\ln(4R/h)}- A \frac{\pi Y \dot
 h}{\ln(4R/h)}\label{ftot},
\end{equation}
 where $A$ is an undetermined constant.
 Figure 2 is the comparison of
our simulation in this model (1189 modes) with the Hertzian contact theory 
(\ref{eq:hertz}).
 The result of our simulation at the impact velocity 
 $v_i=0.01c$ with $c=\sqrt{Y/\rho}$ shows the beautiful hysteresis as 
suggested in the simulation at $v_i=0.1c$. \cite{gerl}
This means the compression and rebound are not symmetric.
The hysteresis curve is still self-similar even at
 $v_i=0.04c$ but the loop becomes noisy at $v_i=0.1c$.

For the low impact velocity $v_i=0.001c$, the hysteresis loop
almost disappears but the total force observed in our simulation is almost a
linear function of $h$ which is a deviation from Hertzian contact theory
and quasi-static theory.
In particular, the turning point at $\dot F=0$ is a deviation
from the Hertzian 
curve. 
This deviation is in clear contrast to the quasi-static theory,
 because the dissipative force in the quasi-static theory
must be zero at the turning 
point which $\dot h=0$ should satisfy.
 In this point, our model seems not to be appropriate
 to represent the nature of slow impact.
 This deviation may be originated from the defect of
 our model. Actually, our model cannot describe the equilibrium
 compression of a disk on the wall.
 We may need thermal deformations which cannot be described
 by elastic deformation of disks.
 The linearity of the total repulsion force is not
 surprising, because
$\textup{e}^{-a y(\phi,t)}$ in the potential term in
eq.(\ref{hamiltonian}) can be expanded in a series of 
$Q_{n,l}$ for very slow impact\cite{gerl,neptis}.

We have numerically studied the impact of a two dimensional
elastic disk with a wall. The results can be summarized as
(i) The coefficient of restitution (COR) decreases with the impact
velocity.
(ii) The result of our simulation is not consistent with the
result of the two-dimensional
quasi-static theory.
 For small velocity, there remains the 
inelastic force even at $\dot h=0$.

We appreciate S. Sasa, Y. Oono, and S. Takesue 
for their valuable comments. One of 
the authors(HK) thanks S. Wada, K. Ichiki, A. Awazu, and
M. Isobe for stimulative discussions. 
This study is partially supported by  
the Grant-in-Aid for Science 
Research Fund from the Ministry of Education, Science and Culture 
(Grant No. 11740228).


\newpage

Fig.1.: Coefficient of restitution for normal collision of our model
 as a function of impact velocity, where
 $c=\sqrt{Y/\rho}$ with the Young's modulus $Y$ and the density $\rho$. 
437 and 1189 modes are chosen for this model.

Fig.2.: The comparison of the Hertzian force in
   eq.(\ref{eq:hertz}) with our simulation at $v_{i}=0.01c$
  (a) and  $v_i=0.001c $(b) at $T=0$ in this model. $F_{tot}$ is the total
  force originated from the wall potential.


\newpage

\newpage
\begin{figure}[thbp]
 \epsfxsize=10cm
 \centerline{\epsfbox{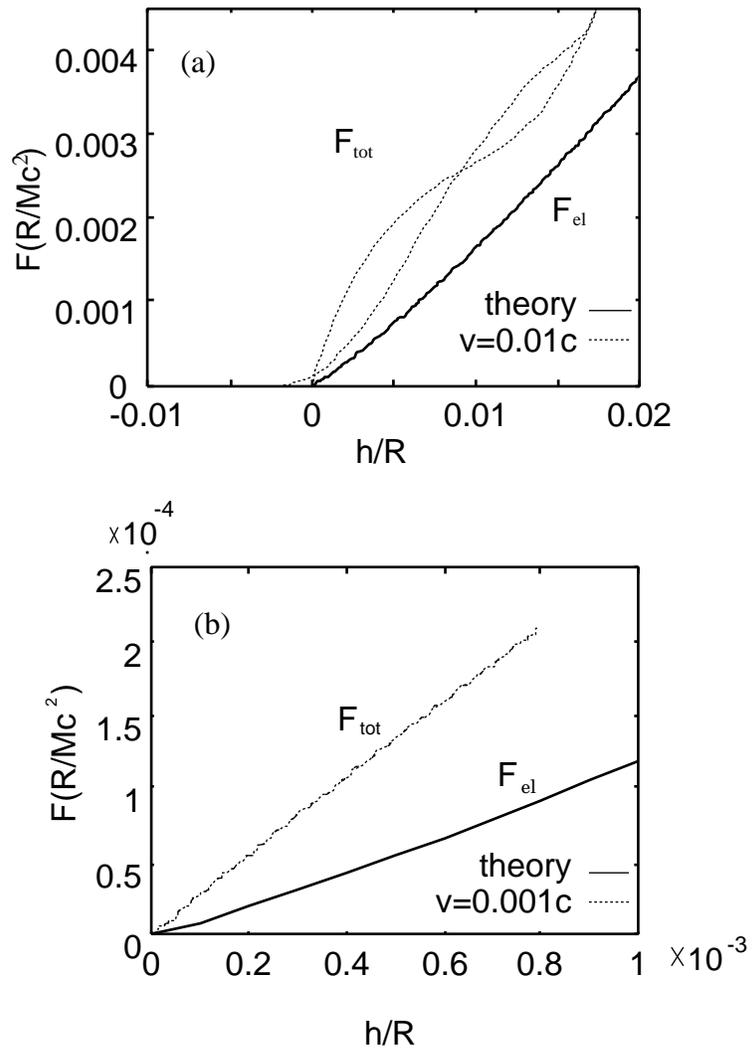}}
 \caption{Authors: Hiroto Kuninaka and Hisao Hayakawa}
 \label{fel2}
\end{figure}

\end{document}